\documentclass[aps,prd,twocolumn]{revtex4-2}

\usepackage[english]{babel} 
\usepackage[utf8]{inputenc}
\usepackage{amsfonts,amsmath,amssymb}
\usepackage[hidelinks,colorlinks=true,linkcolor=blue,citecolor=blue, urlcolor = blue]{hyperref}
\usepackage{hyperref}
\usepackage{graphicx}
\usepackage{multirow}
\usepackage{verbatim}
\usepackage{bm}
\usepackage{paralist}
\usepackage{dcolumn}
\usepackage{latexsym}
\usepackage{natbib}
\usepackage{newunicodechar}
\usepackage{orcidlink}
\usepackage{float}
\usepackage{xcolor}

\newcommand{\ii}{\mathrm{i}}
\newcommand{\ee}{\mathrm{e}}

\newcommand{\ud}{\ensuremath{\mathrm{d}}}

\newcommand{\chisq}{\ensuremath{\chi^{2}}}

\newcommand{\fform}{F}
\newcommand{\bvec}[1]{\ensuremath{\bm #1}}
\newcommand{\dof}{\ensuremath{\text{d.o.f}}}
\newcommand{\apr}{\ensuremath{\alpha^\prime}}

\usepackage{complexity}

\begin{document}
\title{Pomeron Weights in QCD Processes at High Energy and the $S$-Matrix Unitarity Constraint}


\author{Rami Oueslati~\orcidlink{0000-0003-2569-4056}}
\email{rami.oueslati@uliege.be}
\affiliation{Space sciences, Technologies and Astrophysics Research (STAR) Institute, Université de Liège, Bât.~B5a, 4000 Liège,
	Belgium}
\date{\today}
\begin{abstract}
	The pomeron topological cross-section is derived for the eikonal and the $U$-matrix unitarization schemes using a generalized representation of the scattering amplitude at high energy in an effort to examine pomeron characteristics, namely the multiplicity distribution, fluctuation, and correlation, and to reveal the impact of pomeron weights on the $pp$ multiplicity distribution. The results demonstrate that the $U$-matrix inherently incorporates a larger amount of diffraction production into the multi-pomeron vertices, yielding a larger pomerons’ variability regardless of the energy range, while such fluctuations become significant only beyond a specific high-energy threshold in eikonal-like schemes. Most importantly, our findings indicate that within the $U$-matrix scheme, an increase in exchanged pomerons results in more pronounced pomeron correlations and that this correlated resummation plays a key role in enhancing multi-parton collisions. We argue that the choice of unitarization scheme is not arbitrary, as it directly dictates the statistical structure of multi-reggeon exchanges. By reflecting the composite nature of hadrons, rational alternatives to eikonal-like schemes, such as the $U$-matrix, offer a compelling theoretical foundation for modeling high-energy hadronic collisions, provided that specific formulations can be identified that strictly satisfy all mathematical and physical constraints.

	
	
\end{abstract}

\maketitle
\section{INTRODUCTION}

In the study of hadronic interactions at high energy, understanding soft QCD processes that occur 
at low momentum transfer is a daunting task since perturbative QCD techniques are inapplicable and, 
to date, there is no fundamental theory that underlies these processes. That being said, we can approach 
them through phenomenological models that are based on fundamental principles of the $S$-matrix theory, 
such as unitarity, analyticity and crossing symmetry, along with empirical parametrizations. 
In order to improve these models, extensive data comparisons from collider experiments and cosmic-ray 
air showers are necessary to validate their underlying assumptions and fine-tune their parameterizations.

In soft QCD processes, hadronization represents a key phenomenon typically described using 
phenomenological models, such as the Lund string model \cite{ANDERSSON198331}, in combination with 
Gribov–Regge theory \cite{Gribov:1967vfb}. Within these frameworks, unitarization modifies a bare 
scattering amplitude to satisfy the $S$-matrix unitarity constraint, incorporating multi-pomeron 
exchanges primarily through eikonal-like schemes. Inelastic production processes correspond to cut 
pomerons, yielding pairs of color strings between projectile and target partons that fragment into 
final-state hadrons. Thus, pomeron weights directly dictate string formation and configuration. 

While 
such approaches accurately describe a range of hadronic observables, evidence suggests that they remain 
suboptimal for a complete comprehension of the underlying physics, particularly in (ultra-)high-energy 
collisions. 

From a fundamental perspective, it was shown in \cite{Boreskov_2005} that pomeron exchanges in an eikonal-like scheme strictly follow a Poisson distribution. However, since these exchanges occur between composite hadrons consisting of bound quarks and gluons, there is no physical justification for treating them as independent. Pomeron exchanges should instead be interdependent, raising a crucial question regarding the appropriate unitarization scheme to account for such correlations. In this context, directly equating the probability of $n$ string-pair configurations to the probability of exchanging $n$ cut pomerons implicitly assumes that cut pomerons act independently. As a result, the eikonal scheme does not account for potential correlations during string formation \cite{werner1994}.

From a mathematical perspective, rational unitarization schemes, such as the $U$-matrix, are further justified by their strict consistency with $S$-matrix analyticity, unlike exponential schemes, such as the eikonal,  which may yield amplitude structures not fully constrained by analyticity \cite{Blankenbecler}.
From a phenomenological point of view, several studies highlight limitations associated with eikonal-like schemes. For instance, in \cite{Martynov_2020} it was demonstrated that shadow corrections to the rapidly rising contribution of the input supercritical pomeron do not resolve the Finkelstein–Kajantie problem, and it was argued that alternative schemes are necessary. 

A similar situation arises at ultra-high energies, for example in cosmic-ray physics, where the development of extensive air showers is highly sensitive to the properties of the first few high-energy hadronic interactions in the atmosphere. In this regime, such an approximation may break down or introduce biases when extrapolated beyond their validated energy range \cite{Ohashi, Parsons_2011}. 
In response to these limitations, several phenomenological studies have explored alternative unitarization frameworks, notably the $U$-matrix, revealing key differences in their description of the underlying physics. For example, recent work \cite{Oueslati:2024awy} introduced a model relating KNO scaling to geometrical scaling violations, employing the $U$-matrix scheme to describe hadronic multi-particle production. This approach demonstrated that the $U$-matrix offers a viable alternative to the eikonal scheme while raising fundamental questions about the nature of pomeron exchanges under differing unitarization prescriptions, even though both satisfy the unitarity constraint.

Along similar lines, in \cite{Oueslati:2023tjt}, a multi-channel model using the $U$-matrix scheme was proposed to account for more pronounced hadronic fluctuations in diffractive states and to provide a more accurate description of diffractive cross-sections. Remarkably, the insensitivity of the results to the details of the diffractive configurations suggests that the $U$-matrix scheme is incompatible with uncorrelated pomerons exchange.

Overall, multi-pomeron exchange significantly influences high-energy hadron scattering, with their weights substantially affecting hadron amplitudes and consequently directly impacting hadron process cross-sections \cite{Feofilov:2017cle, Bleibel:2010ar}. Owing to the scheme-sensitive nature of the results, this study set out to understand the nature and role of pomeron exchanges in the $U$-matrix scheme in comparison to an eikonal-like one. It also sought to illuminate the impact of the pomeron weights on QCD processes.

This work is organized as follows. In the next section, we will outline the theoretical framework of the pomeron vertices in Gribov-Regge theory, along with a detailed review of the generalized representation of the unitarized elastic scattering amplitude, as proposed by Kancheli. Section III will present and thoroughly discuss the results. Section IV will summarize the main findings and implications of the study.

\section{Pomerons Vertices in Gribov-Regge Theory}\label{sec:pomerons}

Unitarization in the Gribov-Regge theory \cite{Gribov:1967vfb} is accomplished by summing the contributions from all multi-reggeon exchanges. Using this method, the reggeon vertices, referring to the coupling between the exchanged reggeons and the external particle, are used to calculate the amplitudes for multi-reggeon exchanges and their values determine the weights of the n-reggeon exchange. 

It should be emphasized that it is difficult to compute the values of these vertices from first principles. In practice, they are parameters determined phenomenologically by fitting experimental data using some functional form. However, as previously stated in the first section, there is no specific reason to assume that these weights should adhere to the Glauber-eikonal form, suggesting the need for a more comprehensive analysis.

In \cite{Kancheli:2013cra}, Kancheli extended these multi-reggeon vertices to account for more complex interactions, in particular the contribution from diffractive production.  This section will provide a comprehensive review of his formalism, as it establishes the theoretical groundwork essential for our objectives.

We begin with the expression of the unitarized elastic hadronic amplitude represented as follows :

\begin{equation}
	F(s, t) = \sum_{n=1}^{\infty} F_n(s, t),
	\label{eq:elas_ampli}
\end{equation}

with the \( n \) reggeon exchange amplitude given by :
\begin{multline}
	F_n(s, t \simeq -k_{\bot}^2) = \frac{-i}{n n!}\int  N_n^2(k_{\bot i}) \\
	\cdot \prod_{i=1}^n \frac{d^2 k_{\bot i}}{(2\pi)^2} \cdot D(s, k_{\bot i}) \delta^2\left(k_{\bot} - \sum k_{\bot i}\right),
	\label{eq:elas_ampli_n}
\end{multline}
where at high energy, the primary amplitude $F_{1}(s,t)$ can be represented either as a pomeron exchange or as a more intricate set of reggeon diagrams and its factorized form is given by
\begin{equation}
	F_1(s,t) = G(k_{\bot}) D(s, k_{\bot}) G(k_{\bot}), 
	\label{eq:born_ampli}    
\end{equation} where $ D(s, k_{\bot})$ refers to the Green function of the pomeron. As stated in \cite{Gribov_1983}, the vertex function $ N_n(k_{\bot i})$ in (\ref{eq:elas_ampli_n}), representing the emission of $n$ pomerons with transverse momenta $k_{\bot i}$ by the external hadron particle, can be expressed through integrals of the product of $G$ vertices.  Likewise, the vertices can be expanded over on mass shell states of diffractive-like beams \cite{Gribov1968},  thereby accounting for the contribution from diffractive production in the multi-reggeon exchange weights : 
\begin{multline}
	N_n(k_i)= \sum_{\nu_1 ,\nu_2,.. \nu_n}\int G_{1 ~\nu_1}(P_{in},p_i^{(1)})  G_{\nu_1 \nu_2}(p_i^{(1)}, p_j^{(2)}) \cdots  \\
	\cdots  G_{\nu_{n-1} 1}(p_i^{(n-1)} , P_{out}) ~
	\prod_{i=1}^{n-1} d\Omega_{\nu_i}(p_i^{(1)}),
	\label{eq:vertex_expan}
\end{multline} 
where  $G_{\nu_1 \nu_2} (p_i^{(1)}, p_j^{(2)}, k_{\bot})$ is the transition amplitude for a beam of $\nu_1$ particles with momenta $p_i^{(1)}$ into a beam of $\nu_2$ particles with momenta $p_j^{(2)}$, and with the emission of a pomeron with the transverse momentum $ k_{\bot}$.  In (\ref{eq:vertex_expan}), the $d\Omega_{\nu} (p_i)$ represents the element of the $\nu$ particles phase-space volume. The Eq.~(\ref{eq:vertex_expan}) incorporates both the summation and integration over all conceivable physical states of the particles in the beams and thus accounts for their full masses. It is important to note that the multi-particle amplitudes $G_{\nu_{1} \nu_{2}}$ are complex and may contain unrelated contributions, whereas the vertex functions $N_n (k_i)$ are real.

Considering a non-local field operator $\hat{G}(k)$  describing the pomeron emission vertices $G_{\nu_1 \nu_2} (k)$ between the initial and final states of the external particle, the expression (\ref{eq:vertex_expan}) for the vertex functions $N_n (k_i)$ can be written in a symbolic operator form as the average of the product of this field operator $\hat{G}(k)$ : 

\begin{equation}
	N_n(k_i) = \langle P_{in}|\hat G(k_1) \hat G(k_2) \cdots \hat G(k_n)|P_{out}\rangle 
	\label{eq:vertex_op}
\end{equation} This product can further be decomposed over the complete set of physical states of the beams $|\nu \rangle $ as follows  
\begin{widetext}
	\begin{equation}
		N_n(k_i) = \sum_{\nu_1,\dots,\nu_{n-1}} \langle P_{in}|\hat G(k_1)|\nu_1\rangle \langle \nu_1 |\hat G(k_2)|\nu_2\rangle \langle \nu_2 |\hat G(k_3)|\nu_3 \rangle \cdots \langle \nu_{n-2} |\hat G(k_{n-1})|\nu_{n-1}\rangle 
		\langle \nu_{n-1} |\hat G(k_n)|P_{out}\rangle
		\label{eq:vertex_op}
	\end{equation}
\end{widetext} One can simplify the handling of the vertex operators $ \hat{G}(k)$ by redefining the basis for the beam states $|\nu \rangle$ where $\hat{G}(k)$ has a simple diagonal form : 

\begin{equation}
	\hat{G}(k) | \nu \rangle = g_{\nu}(k) | \nu \rangle, 
\end{equation}  
with $g_{\nu}(k)$ acting as eigenvalues. Then the expression Eq.~(\ref{eq:vertex_op}) is simplified as the summation over all possible beam states, with the contribution from each state weighted by a function $w(\nu)$ and the product of its associated vertex functions for the different momentum components $k_{i}$ :
\begin{equation}
	N_n(k_i) ~= \sum_{\nu} w(\nu)~
	\prod_1^n g_{\nu}(k_i)
	\label{eq:vertex_diag}
\end{equation}
where  $w(\nu) $ can be interpreted as the probability of finding the fast hadron in the state $| ~\nu \rangle$ and is given by:
\begin{equation}
	w(\nu) = \langle  P_{in} |\nu \rangle~
	\langle \nu |  P_{out} \rangle  
\end{equation} Following this simplification, the $S$-Matrix in the impact parameter representation can be written as follows: 
\begin{eqnarray}
	S(s,b) &=&  \sum_{\nu_1 \nu_2} w(\nu_1) w(\nu_2)~\sum_{n=0}^{\infty} \frac{(~i~ \chi_{\nu_1 \nu_2} (s,b))^n}{n!}\\ \nonumber
	&=& \sum_{\nu_1 \nu_2} w(\nu_1) w(\nu_2) e^{i \chi_{\nu_1 \nu_2}(s,b)}
\end{eqnarray}
where
\begin{equation}
	\chi_{\nu_1 \nu_2}(s,b) = \int d^2 k_{\bot}e^{i b k_{\bot} } g_{\nu_1}(k_{\bot}) D(s,k_{\bot}) g_{\nu_2}(k_{\bot})    
\end{equation}
One can further simplify the expression for the vertices by factorizing the vertex function $g_{\nu}(k)$ into a universal term $g(k)$ and a state-dependent coefficient $\lambda(\nu)$, with a small non-factorizable correction $\tilde{g}(\nu, k)$ : 
\begin{equation}
	g_{\nu}(k) ~=~ g(k) \lambda(\nu) + \tilde{g}(\nu, k)  
\end{equation} then the vertices can be represented as :
\begin{equation}
	N_n(k_i) ~\simeq~ \beta_n~ \prod_1^n g(k_i),
	\label{eq:vertex_facto}
\end{equation} where
\begin{equation}
	\beta_n  ~=~   \sum_{\nu} w(\nu)~(\lambda(\nu))^n
	\label{eq:beta_n_moment}
\end{equation}
Under this assumption of factorization, the vertices simplify to a product of universal functions $g(k_i)$, weighted by $\beta_n$, which encapsulates the sum over the probabilities $w(\nu)$ and the coefficients 
$(\lambda(\nu))^n$. It is worth noting that within the Glauber-eikonal approximation, where the single-particle state in beams contributes very little to $N_n(k_i)$, the integrals in (\ref{eq:vertex_expan}) disappear. In this case, the $n$-pomeron vertex factorizes into a product of elastic pomeron couplings, where the expression for $N_n(k_i)$ becomes a simple product of $n$ elastic pomeron vertices as :

\begin{equation}
	N_n(k_i) = \prod_{i=1}^n g(k_i),
	\label{eq:vertex_eiko}
\end{equation} 
where $g(k)=G_{11}(p,p+k)$

All in all we obtain the following expression for the $S$-matrix in the impact parameter representation :

\begin{equation}
	S(s,b) = \sum_{n=0}^{\infty} \frac{\beta_n^2}{n!} \left( i \, \chi(s,b) \right)^n
	\label{eq:generalized_smatrix}
\end{equation} 
where the real coefficients $\beta_n \geq 1$ are not fixed a priori, but will be determined depending on the unitarization scheme chosen for the elastic amplitude, which will be illustrated in the forthcoming section. Eq.(\ref{eq:generalized_smatrix}) represents a generalized expansion of the $S$-matrix, where the function $\chi(s,b)$ encodes the underlying scattering dynamics, while the coefficients $\beta_n$ control the relative weight of the different orders of the interaction terms.

Series expansions of the form Eq.(\ref{eq:generalized_smatrix}), where the multi-reegeon vertex depends only on the number of particles exchanged, are common in the literature \cite{martynov2014} and encompass various unitarization schemes for appropriate choices of the coefficients $\beta_n$. However, working directly with an infinite series is often impractical and can obscure its convergence properties. Yet, it might be possible to gain a better understanding of the series' structure by introducing a spectral representation of the coefficients. 


Since the coefficients $\beta_n$ in (\ref{eq:beta_n_moment}) are defined by summing factors raised to the $n$th power with their respective weights, they can naturally be interpreted as the moments of a spectral distribution. 
In the limit of a very large number of contributing states, the discrete set ${\lambda(\nu)}$ with corresponding weights $w(\nu)$ can be mapped to values of a continuous variable $\tau$ and an associated spectral function $\varphi(\tau)$, allowing the sum to be expressed in the continuum limit as :

\begin{equation}
	\beta_n = \int_0^{\infty} d \tau \, \tau^n \, \varphi(\tau)~,
\end{equation}
then the $S[\chi]$ matrix can be rewritten as a combination of Glauber-type eikonal terms:

\begin{equation}
\label{eq:gene_smat}
S(s,b)\equiv S[\chi] = \int_0^{\infty} d\tau \, \rho(\tau) \, e^{i\tau \chi(s,b)}
\end{equation}

with
\begin{equation}
\rho(\tau) = \int_0^{\infty} \frac{d\tau_1}{\tau_1} \, \varphi(\tau_1) \, \varphi(\tau/\tau_1)
\end{equation} functions as a weight. The constraints $\beta_0 = \beta_1 = 1$ are imposed by the normalization condition for $S[\chi]$ and $w(\nu)$. This results in the following relations:
\begin{equation}
\label{norm_beta}
\int_0^{\infty} d\tau \, \rho(\tau) = \int_0^{\infty} d\tau \, \tau \rho(\tau) = 1
\end{equation}
Using the generalized $S$-matrix representation (\ref{eq:gene_smat}), we can write some general relationships for cross-sections at a given impact parameter, which are valid for any spectral density $\rho(\tau)$ :

\begin{itemize}
	\item The total cross-section :
		
	\begin{equation}
		\label{{totcs}}
		\sigma_{tot}(s,b) = 2~(1 - Re(S[\chi])),
	\end{equation}  
	
	\item The elastic cross-section
	
	\begin{equation}
		\sigma_{el}(s,b) = |~1 - S[\chi]~|^2,
	\end{equation} 
	
	\item The total inelastic cross-section
	
	\begin{equation}
		\sigma_{in}(s,b) = \sigma_{tot} -  \sigma_{el}  =
		1 - |S[\chi]|^2~~~
	\end{equation}
	\item The total cross-section of diffraction generation : single $\sigma_{sd}$ and double $\sigma_{dd}$
	
	\begin{align}
		\sigma_{dif}(s,b) &=  \sigma_{in} - \sigma_{\text{in, cut}} \nonumber
		=    2 \sigma_{sd} + \sigma_{dd} \\
		&= S[2 i Im(\chi)] -|~S[\chi]~|~^2~~
		\label{difgen}
	\end{align}
	\item The cross-section corresponding to processes when at least one pomeron is s-cut : when we cut a single pomeron from the elastic amplitude, this corresponds to taking the imaginary part of the corresponding partial-wave amplitude. From the optical theorem, we know that this is related to the cross-section for cutting a single pomeron. Therefore 
	
	\begin{equation}
		\sigma_{\text{in, cut}}(s,b) = 1 - S[2 i ~Im (\chi)]
	\end{equation} 
	and so the pomeron topological cross-section: these are the contributions of diagrams with $n$ cut pomerons and of the arbitrary number of uncut pomeron lines
	
	\begin{equation}
		\sigma_n(s,b) = \int_0^{\infty} d\tau \rho(\tau)
		~\frac{(2\tau Im(\chi))^n}{n!} ~e^{-2 \tau Im(\chi)}  
		\label{eq:sig_ncut}
	\end{equation}  This expression resembles a superposition of Poisson distributions, where each term represents the contribution of a Poisson distribution with mean $(2\tau \text{Im}(\chi))$ and weighted by the spectral density $\rho(\tau)$.
	
\end{itemize}

\section{RESULTS}

\subsection{Pomeron topological cross-section}

In this section, we use the generalized spectral representation of the impact-parameter $S$-matrix to illustrate how multi-pomeron dynamics are embedded in different unitarization schemes. Within this framework, using Eq.~(\ref{eq:generalized_smatrix}), and the constraint $\beta_0 =  1$, the generalized expansion of the unitarized elastic amplitude in impact parameter space reads :

\begin{equation}
	A(s,b) = - i ~ \sum_{n=1}^{\infty} \frac{\beta_n^2}{n!} \left( i \, \chi(s,b) \right)^n
	\label{eq:general_elastic}
\end{equation}
This amplitude is entirely governed by the choice of the spectral function $\rho(\tau)$, or equivalently by the coefficients $\beta_n$ subject to the constraints. Choosing a trivial spectral density concentrated at a single value, 
$\rho(\tau) ~=~ \delta (\tau - 1) ~,$ the corresponding unitarized elastic amplitude reads

\begin{equation}
	A(s, b) = \ii \left[ 1 - \ee^{\ii \chi(s, b)} \right]\,
	\label{eq:eikx}
\end{equation} which is the standard eikonal expression \cite{Cudell:2008yb}. In this case, the pomeron topological cross-section follows directly from Eq.~(\ref{eq:sig_ncut}) :
\begin{equation}
	\sigma_n(s, b) = \frac{(2\text{Im}(\chi))^n}{n!} e^{-2\text{Im}(\chi)}
	\label{eq:sig_n_eiko}
\end{equation}
A different realization is obtained by adopting an exponential spectral density,
\begin{equation}
	\rho(\tau) = 4 ~\mathrm{e}^{- 2 ~\tau} - \delta(\tau)
\end{equation} for which the elastic amplitude becomes
\begin{equation}
	A(s, b) = \frac{\chi(s, b)}{1 - \ii \chi(s, b)/2}
	\label{eq:umatx}
\end{equation} which is the $U$-Matrix form of the unitarization scheme  \cite{Cudell:2008yb}. The associated pomeron topological cross-section follows from Eq.~(\ref{eq:sig_ncut}) and is given by : 
\begin{equation}
	\sigma_n(s,b) = 2~\frac{(\text{Im}(\chi))^n}{(1 + \text{Im}(\chi))^{1+n}}
	\label{eq:pom_topo_umat}
\end{equation}

Notably, the two unitarization schemes lead to qualitatively different expressions for the pomeron topological cross-section. This reflects the fact that distinct schemes and hence different implementations of the unitarity constraint arise naturally from the generalized $S$-matrix representation through the choice of the spectral density. A direct manifestation of these differences appears in the probabilistic interpretation of multi-pomeron exchange. While in the eikonal case the number of exchanged pomerons is a random variable Poisson distributed in impact parameter space \cite{Boreskov_2005}, it is worth investigating how this distribution is modified in alternative frameworks, most notably within the $U$-matrix approach. To do so, observing that the expression (\ref{eq:sig_ncut}) of the pomeron topological cross-section $\sigma_n(s,b)$, can be understood as a mixed Poisson distribution, in which the conditional distribution of the number of pomeron exchanged, given a certain rate parameter, is Poissonian. Nevertheless, the rate parameter itself is handled as a random variable with its own probability distribution \cite{johnson1992univariate}.  Thus, one can question what kind of physical features can be obtained with this random rate parameter rather than with a fixed Poisson rate parameter for all events. 
To achieve this,  let $X$ be a random variable following a mixed Poisson distribution with density $\pi(\lambda)$. Its probability mass function is given by : 
\begin{equation}
	Pr(X=n)=\int _{0}^{\infty }{\frac {\lambda ^{n}}{n!}}e^{-\lambda }\,\,\pi (\lambda )\,\mathrm {d} \lambda.    
\end{equation} Additionally considering that the Poisson rate parameter is distributed according to an exponential distribution, $\pi(\lambda )={\frac {1}{\gamma }}e^{-{\frac {\lambda }{\gamma }}} $  and using integration by parts n times yields:

\begin{equation} 
	Pr(X=n)= \left({\frac {\gamma }{1 + \gamma }}\right)^{n}\left({\frac {1}{1+\gamma }}\right)
\end{equation} Accordingly when a Poisson distribution is compounded with an exponentially distributed rate parameter, the resulting distribution is geometric,
\begin{equation}
	X \sim \operatorname{Geo}\!\left(\frac{1}{1+\gamma}\right).
\end{equation} This structure is realized explicitly in the $U$-matrix unitarization scheme. Indeed, using Eq.~(\ref{eq:pom_topo_umat}), the inelastic overlap function is obtained by summing over all cut multiplicities :

\begin{equation}
	G_{\rm inel}(s, b) = \sum_{n=1}^{\infty} \sigma_n(s, b)
	= \frac{2\, r}{\bigl[1 + \operatorname{Im} \chi(s, b)\bigr] (1 - r)}
	\label{eq:Ginel_umat}
\end{equation} where
\begin{equation}
	r = \frac{\operatorname{Im} \chi(s,b)}{1 + \operatorname{Im} \chi(s,b)} < 1
\end{equation} and so the probability of having exactly n cut pomerons at a fixed impact parameter b is given by : 

\begin{equation}
\label{eq:Prob_succes}
P_n(s, b) = \frac{\sigma_n(s, b)}{G_{\rm inel}(s, b)} 
= p \, (1-p)^{\,n-1} 
\end{equation} where
\begin{equation}
\label{eq:prob_succes}
p = \frac{1}{1 + \operatorname{Im} \chi(s,b)}
\end{equation} Consequently,  within the $U$-matrix scheme, the number of pomerons is a random variable that follows a geometric distribution. This distribution arises from a Poisson process with a fluctuating rate parameter and reflects the fact that pomeron exchanges are no longer statistically independent.  This dependency implies collective phenomena at fixed impact parameter, such as correlation among the exchanged pomerons. This  feature sharply distinguishes the $U$-matrix scheme from others, particularly the eikonal, which lacks these properties.

Interestingly, our results in (\ref{eq:pom_topo_umat}) and (\ref{eq:sig_n_eiko}) coincide with those of \cite{Luna:2024zoq}, where the pomeron topological cross-section is derived by applying the AGK cutting rules and using the coefficients obtained in each unitarization scheme through a power-series expansion of the elastic scattering amplitude in impact-parameter space in terms of the Born amplitude. More remarkably, both formalisms imply a significant mathematical contradiction at very high energy in the case of the $U$-matrix scheme :  
	
	On the one hand, the unitarity equation indicates that the inelastic overlap function $G_{\text{inel}}$ approaches $0$:
	\begin{align}
		\label{uamp}
		G_{\text{inel}}(s,b)
		&= 2\,\mathrm{Im}\,\mathcal{A}(s,b) - |\mathcal{A}(s,b)|^2 \nonumber \\
		&= \frac{2\,\mathrm{Im}\,\chi(s,b)}
		{\left(1 - i\,\chi(s,b)/2\right)\left(1 + i\,\chi^{*}(s,b)/2\right)}
		\;\xrightarrow{\chi \to \infty}\; 0.
	\end{align}  
	
	On the other hand, the sum of the cross-sections according to the AGK cutting rules or the spectral representation approaches $2$:
	\begin{equation}
		G_{\mathrm{inel}}^{(\mathrm{AGK}/\text{this formalism})} \to 2.
	\end{equation} This mathematical inconsistency is explained by the nature of the spectral density $\rho(\tau)$ in the $U$-matrix case. It contains a negative term, meaning it does not represent a true classical probability density, but rather quantum  effects. Physically, this corresponds to the reflective scattering mode. At small impact parameters, the hadron core no longer behaves as a perfect absorber, but as a reflector. The elastic wave is reflected, and particle production vanishes at the center ($G_{\text{inel}} \to 0$), forming a ``black ring'' at the periphery \cite{Troshin_2019}. It should be noted that, if the spectral density $\rho(\tau)$ is strictly non-negative on $[0, \infty)$, a required condition to preserve probability in the AGK sense, the $S$-matrix for a purely absorptive born amplitude ($\chi = i\Omega$) is given by:
	\begin{equation}
		S(\Omega) = \int_0^\infty \rho(\tau) e^{-\tau \Omega} d\tau > 0 \quad \forall \Omega \ge 0.
	\end{equation} Consequently, no model with a strictly non-negative spectral density $\rho(\tau) \ge 0$ can generate reflective scattering mode, as occurs in the eikonal scheme.
	Owing to the ambiguity in constructing scattering amplitudes at high energy in the non-perturbative regime of QCD, one might instead employ an $S$-matrix bootstrap approach \cite{Correia:2025uvc}. This method restricts the space of acceptable spectral densities by imposing physical and mathematical constraints. In this framework, different unitarization schemes represent specific points in this space, the validity of which can be evaluated against experimental data.

\subsection{Pomeron multiplicity distribution}

To reflect the distinct physical behaviors associated with the pomeron weights in both schemes, let us consider an explicit model for $p+p(\bar{p})$ collisions at high energy. The starting point is the single pomeron scattering amplitude, i.e. the Born term, we parameterise it as

\begin{equation}
	a(s,t) =g_p^2\, \fform_1(t)^2 \left( \frac{s}{s_0} \right)^{\alpha(t)}\, \xi(t)\,,
	\label{eq:amp_t}
\end{equation} using the pomeron trajectory $\alpha(t)$,  the proton elastic form factor $F_1(t)$ and the coupling pomeron-proton $g_p$, with $\xi(t)$ the signature factor
\begin{equation}
	\xi(t) =-e^{-i\pi\alpha(t)\over 2}\,.
	\label{eq:amp_t2}
\end{equation} and the pomeron trajectory close to a straight line  \begin{equation}
	\alpha(t)=1+\epsilon+\apr t.
	\label{eq:amp_t3}
\end{equation} In the impact-parameter representation, where the Fourier transform of the amplitude $a\left( s, t \right)$ rescaled by $2s$ is equivalent to a partial wave
\begin{equation}
	\chi(s, \bvec{b}) = \int \frac{\ud^2\bvec{q}}{\left( 2\pi \right)^2}
	\frac{a(s,t)}{2s} \ee^{\ii \bvec{q}\cdot \bvec{b}}.
	\label{eq:Gsb}
\end{equation} We used a dipole-like form factor for the proton $F_1 = 1/(1-t/t_0)^2$. The parameters  $\epsilon$ and $\alpha^\prime$ describing the pomeron trajectory, the coupling constant $g_p$ and $t_0$  the form-factor scale, are adjusted from a fit to up-to-date hadron collider data on total, elastic and inelastic cross-sections both for the eikonal and $U$-matrix unitarisation schemes \cite{Bhattacharya:2020lac} and are provided in table \ref{tab:fits1}.
\begin{table*}
	\begin{tabular}{l@{\hspace{1.5em}}c@{\hspace{1.5em}}c@{\hspace{1.5em}}c@{\hspace{1.5em}}c@{\hspace{1.5em}}r}
   \hline
   \hline\\[-1em]
   Scheme   & $\epsilon $
            & $\apr $
            & $ g_p $
            & $ t_{0} $
            & $ \frac{\chisq}{\dof} $\\[0.3em]
            \hline\\[-0.6em]
   Eikonal
            & $ 0.11 \pm 0.01 $
            & $ 0.31 \pm 0.19 $
            & $ 7.3  \pm 0.9  $
            & $ 1.9  \pm 0.4  $
            & $ 1.442 $ \\[0.5em]
   U-matrix
            & $ 0.10 \pm 0.01 $
            & $ 0.37 \pm 0.28 $
            & $ 7.5  \pm 0.8 $
            & $ 2.5  \pm 0.6  $
            & $ 1.436 $ \\[0.5em]
   
   \hline
   \hline
\end{tabular}

	\caption{\label{tab:fits1}\chisq/\dof\ and best-fit parameters obtained using the eikonal and $U$-matrix unitarisation schemes.}
\end{table*}

\begin{figure}[htbp!]
	
	\includegraphics[width=0.5\textwidth]{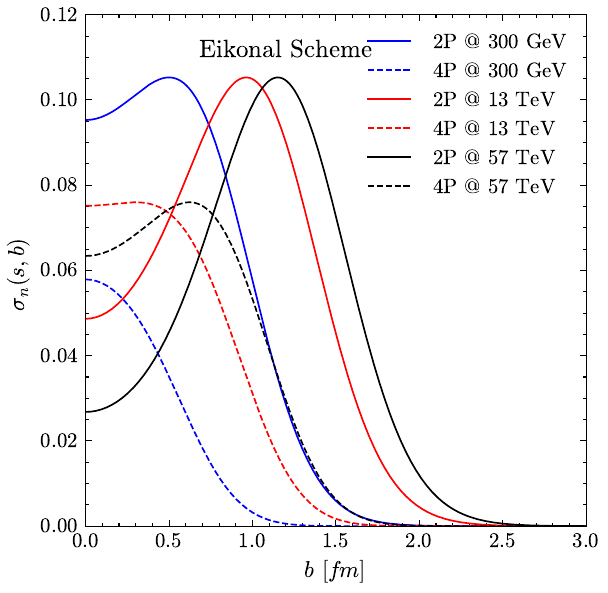}
	\caption{\label{fig:b_space_pom_xsec_eiko} Impact parameter evolution of the Pomeron Topological cross-section in the Eikonal case.}
\end{figure}

\begin{figure}[htbp!]
	\includegraphics[width=0.5\textwidth]{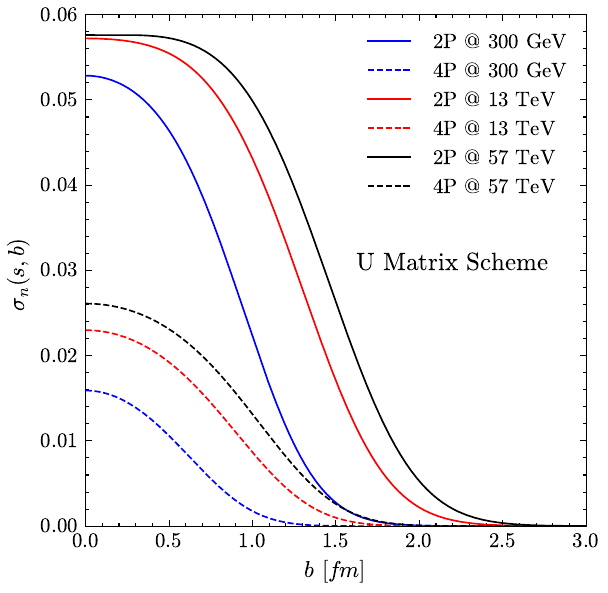}
	\caption{\label{fig:b_space_pom_xsec_umat} Impact parameter evolution of the Pomeron Topological cross-section in the $U$-Matrix case.}
\end{figure}

In order to understand the hadronic dynamics at high and ultra-high energies, particularly in terms of the spatial distribution of the interactions and their implications for particle production, we examined the behaviour of the pomeron topological cross-section in the impact parameter space with energy and the number of pomerons exchanged, in both the eikonal and $U$-matrix cases, as shown in Fig.~\ref{fig:b_space_pom_xsec_eiko} and Fig.~\ref{fig:b_space_pom_xsec_umat}. 

As can be seen from both figures, this function exhibits a distinct pattern in impact parameter space regardless of the scheme used. Nevertheless in each scheme, it has a broadly similar shape irrespective of both the energy level and the number of the pomerons exchanged. More specifically, as shown in the eikonal case, this function is predominantly peripheral, indicating a substantial contribution at large impact parameters. This implies that the majority of pomeron interactions are more likely to occur when the colliding hadrons pass through each other at large distances, reflecting that the interactions are "softer" in nature, meaning they involve long-range processes, likely mediated by soft pomeron exchanges. 
On the other hand, in the $U$ matrix case, we can clearly see that this function is primarily central, suggesting that pomeron interactions tend to happen when the colliding hadrons pass through each other in close proximity, entailing that the collisions tend to involve higher energy densities, leading to more intense interactions in the core of the colliding hadrons.

Furthermore, another aspect observed from these figures is that, with the exponential scheme, this function tends to decline in the same manner with respect to the impact parameter as energy increases. Notably, for a given $n$,  the peak of this function shifts toward larger values of $b$ with increasing energy, while the magnitude of the peak remains approximately constant. This suggests that the spatial region, where the interactions take place, expands with increasing energy, making peripheral collisions even more dominant at ultra-high energy. Most importantly, this peripheral behaviour and constancy of the peak's magnitude are indicative of reaching a saturation effect where the available parton density limits the increase in the interaction strength even with increasing energy.

Whereas, with the rational scheme, this cross-section typically shows a more gradual decrease with increasing impact parameter as energy rises.  More interestingly, for a given $n$, the maximum remains near the centre of the collision, and its value increases with energy. This indicates that the strength of the pomeron interactions grows at the core of the collision and we can understand that more partons are involved in the interaction in the central region.  This reflects a regime where the interactions are still increasing, indicating that saturation has not yet been reached in the central collision region. 

Additionally, according to these figures, it is clear that,  for a fixed energy,  with two and four pomerons exchanged the magnitude of this function decreases as $n$ increases for both schemes. We may suggest that higher-order pomeron exchanges become less significant at higher energies depending on their nature.

\begin{figure}[htbp!]
	\includegraphics[width=0.5\textwidth]{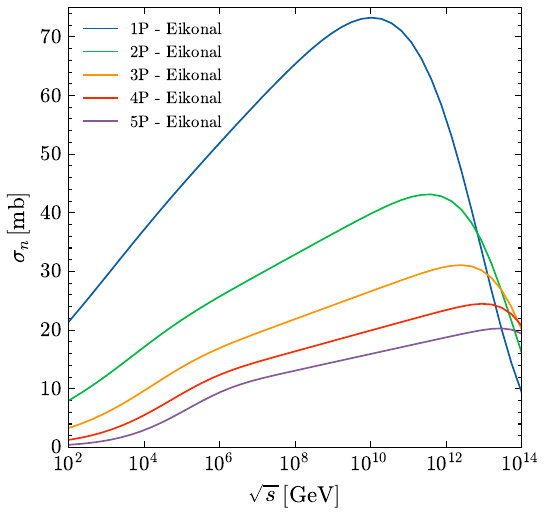}
	\caption{\label{fig:energy_evolution_pom_xsec_eiko} Energy evolution of the Pomeron topological cross-section in the Eikonal case.}
\end{figure}

\begin{figure}[htbp!]
	\includegraphics[width=0.5\textwidth]{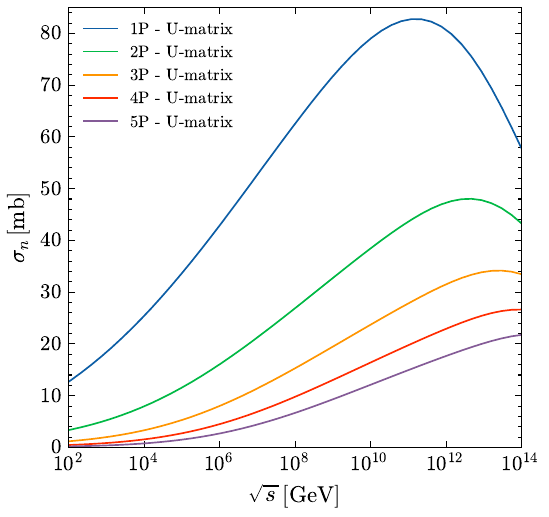}
	\caption{\label{fig:energy_evolution_pom_xsec_umat} Energy evolution of the Pomeron topological cross-section in the $U$ Matrix case.}
\end{figure}

The energy evolution of the pomeron topological cross-sections for $1,2, 3, 4$ and $5$-pomerons exchanged has also been examined in both the eikonal and $U$-matrix cases. By looking at Fig.~\ref{fig:energy_evolution_pom_xsec_eiko} and Fig.~\ref{fig:energy_evolution_pom_xsec_umat}, we can generally see that these cross-sections roughly exhibit the same behavior in both schemes.
To be more specific, for each pomeron exchanged, the cross-sections increase as energy rises, with the contribution of $1$-pomeron exchanged showing the highest value across all energy ranges.  This suggests that the interactions are primarily governed by the simplest diagrams in the pomeron exchange framework, especially at lower energies. As for the higher-order pomeron exchanges $\sigma_{2pom}$, $\sigma_{3pom}$, etc.), they contribute gradually less. It also shows that all cross-sections tend to reach a maximum then decline abruptly at extreme energies, indicating a signature of the unitarity constraint. It is significant to note that in comparison with the eikonal case, this unitarity constraint signature is hit at a slightly higher energy for each pomeron exchanged in the $U$-matrix case. Indeed, this demonstrates the different energy levels at which unitarity effects take over for each pomeron contribution in these two schemes.

Furthermore, before reaching the energy threshold of the unitarity signature, it is evident that all cross-sections’ curvatures, starting from $2$ pomerons exchanged, significantly change at energies beyond $10^4$ GeV in the eikonal case. It should be noted that with more pomerons being exchanged, this effect intensifies. Conversely, in the $U$-matrix case, despite taking into consideration several pomeron exchanges, the cross-sections show a more constant and progressive behaviour without any change in curvature.


The pomeron multiplicity distribution $W_{n}(s)$, i.e., the probability of n pomerons exchanged in an inelastic collision at the energy $s$, is given by 
\begin{equation}
	W_{n} =  \frac{\sigma_{n}}{\sum_{n'} \sigma_{n'}} 
\end{equation}

\begin{figure}[htb]
	
	\includegraphics[width=0.5\textwidth]{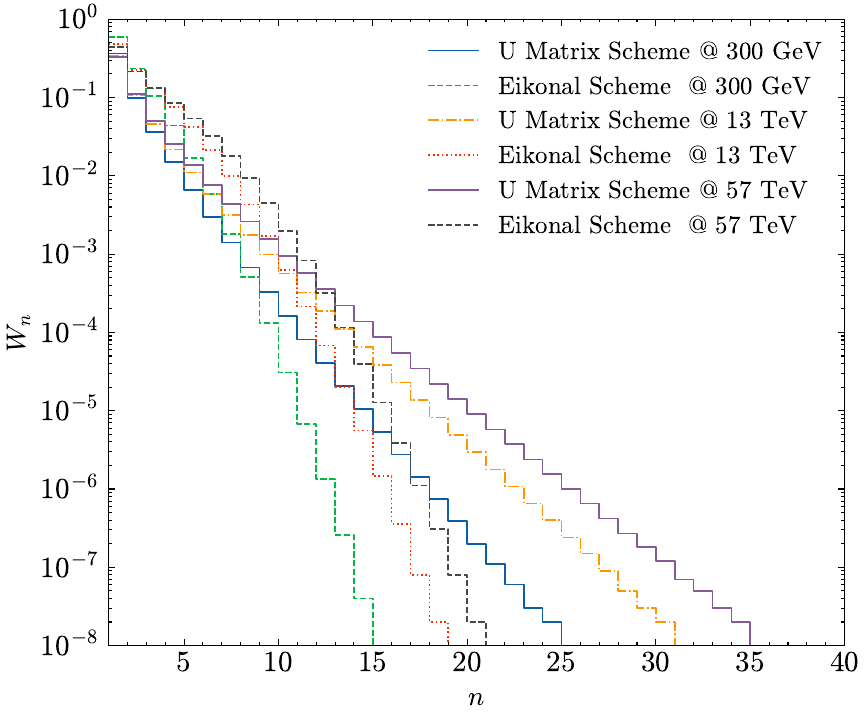}
	\caption{\label{fig:wg_pom_lin} Pomeron multiplicity distribution in both cases, eikonal and $U$-matrix.}
\end{figure}

Using the ansatz for the single pomeron exchange amplitude in the eikonal and $U$-Matrix cases, we compute $W_{n}$. The results are plotted in Fig.~\ref{fig:wg_pom_lin} for three collision energy scales.



Fig.~\ref{fig:wg_pom_lin} clearly shows that in the eikonal case, the exchange of a large number of pomerons is significantly suppressed compared to the $U$-matrix case. In the latter scheme, the exchange of one pomeron enhances the probability of exchanging additional pomerons, and then the pomeron multiplicity distribution would deviate from a Poissonian one. This remarkable difference entails that multi-pomeron exchange is different in the two schemes. In particular, it may result from the presence of collective phenomena, such as correlations between the exchanged pomerons in contrast to what would be expected from an independent exchange. 



It goes without saying that the role of the multi-pomeron exchange becomes more significant as energy grows. In the case of a Poisson distribution, the mean and the variance are equal. In hadronic interactions, however, their relationship tends to vary depending on a number of  factors, such as the energy of the interaction and the impact parameter. As a matter of fact, in order to highlight the role of the multi-pomeron exchange, particularly at high energy, we investigated the energy evolution of the mean and the variance of the number of pomerons exchanged.  Using the probabilities in each scheme, we can calculate the mean and variance of the number of cut-pomerons as a function of the energy:
\begin{equation}
	\langle n \rangle = \sum\limits_{n'=0}^{\infty} n'  W_{n'} 
\end{equation}
and 
\begin{equation}
	\text{Var}(n) = \sum\limits_{n'=0}^{\infty} {n'}^2  W_{n'} - \langle n \rangle^2  .
\end{equation}

\begin{figure}[htb]
	\centering
	\includegraphics[width=0.48\textwidth]{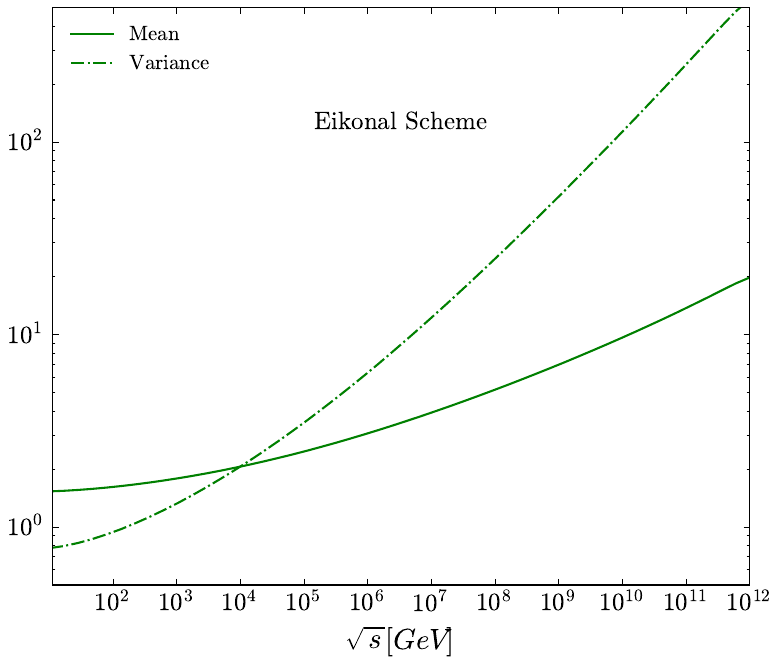}
	\caption{\label{fig:mean_var_eiko} Mean and variance of the number of pomerons
		in the eikonal and $U$ Matrix case}
\end{figure}

By looking at Fig.~\ref{fig:mean_var_eiko}, we can clearly see that in the eikonal case both the mean and the variance of the exchanged pomerons increase more considerably with energy. In addition, it is evident that the mean-variance relationship shows a noticeable shift around  $10^4$ GeV. More precisely, at energies below $10^4$ GeV, we can notice that the variance is steadily smaller than the mean. This suggests that the fluctuations of the exchanged pomerons are not as severe as they are in the average behavior. This also points to a more consistent and stable interaction dynamics at lower energy, where more predictable contributions govern the pomeron exchanges. Nevertheless, at energies above $10^4$ GeV, the variance exceeds the mean with a disproportionate growth.

It is significant to note that the result regarding the energy evolution of both the mean and the variance of the pomerons exchanged aligns with that reported in the quasi-eikonal framework \cite{Bodnia:2013nba}. Most importantly, they uncover a significant distinction in the energy shift of the mean-variance relationship. Indeed, the transition in the quasi-eikonal case, where the variance exceeds the mean, takes place at a lower energy scale, around $200$ GeV. In contrast, this transition is seen at a greater energy, roughly $10^4$ GeV, in the eikonal case. One of the possible explanations for this discrepancy is linked to the fact that the quasi-eikonal unitarization is an extended eikonal-like scheme, with an extra factor $c$ in the expression of this unitarisation scheme accounting for the modification of the simple eikonal due to inelastic diffractive states. For instance, a value of $c = 1.5$ is utilized, corresponding to a $50\% $ contribution of low-mass diffractive states, in comparison to the elastic ones.  Hence, the additional dynamics from the diffractive process in the quasi-eikonal scheme make the number of exchanged pomerons more variable, which lowers the energy threshold for the transition regime in which the variance is greater than the mean. 

\begin{figure}[htb]
	\centering
	\includegraphics[width=0.48\textwidth]{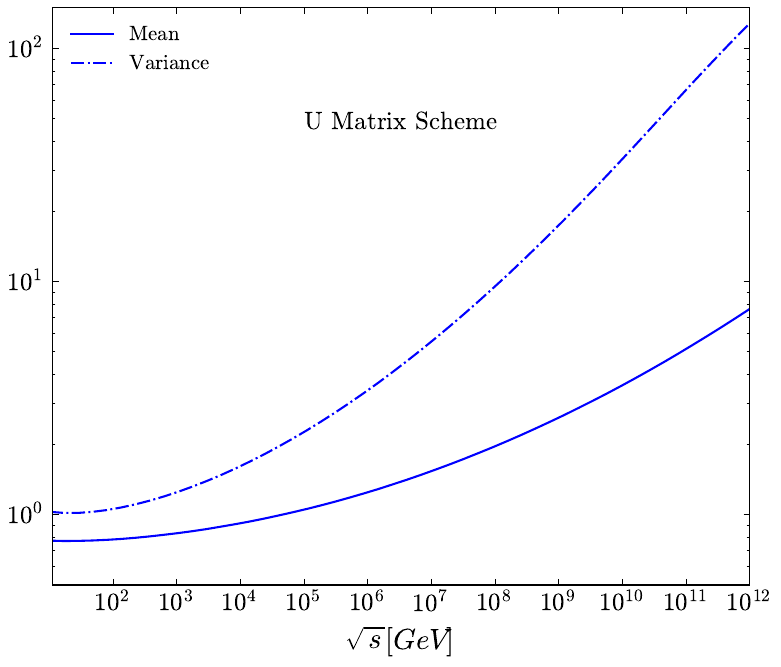}
	\caption{\label{fig:mean_var_umat} Mean and variance of the number of pomerons
		in the eikonal and $U$ Matrix case}
\end{figure}

Fig.~\ref{fig:mean_var_umat} demonstrates the evolution of both the mean and the variance with energy in the $U$-matrix scheme. As can be seen, the mean number of pomerons exchanged progressively rises as energy grows and it keeps increasing at higher energies.

The variance exhibits a similar increase. Surprisingly, however, it remains strictly greater than the mean across the entire energy range, with this overdispersion becoming increasingly pronounced as the energy rises.

As a matter of fact, this striking result can be explained by the structure of the vertices pertaining to hadronic interactions at high energy, with its physical origin likely rooted in the overlap of diffractive states. While the difference between the two schemes regarding the parametrized macroscopic hadronic overlap function is marginal (\ref{eq:amp_t}), their internal dynamics diverge significantly. By expanding the vertices into diffractive beam states (\ref{eq:vertex_facto}), the $U$-matrix scheme intrinsically incorporates a considerable amount of this overlap. This structure reflects a wider range of interaction possibilities, increasing pomeron variability and driving these more pronounced fluctuations. Conversely, the eikonal scheme assumes a simpler vertex structure (\ref{eq:vertex_eiko}) that neglects such overlap. As a result, it exhibits simpler dynamics with restricted variability, and its fluctuations only exceed the mean at much higher energies where multi-pomeron exchanges become substantial.

The larger fluctuations found in the $U$-matrix scheme suggest a more complex interaction dynamics, with a stronger contribution from diffractive processes than in the eikonal and quasi-eikonal approaches. This may explain the results reported in \cite{Vanthieghem:2021akb}, where the $U$-matrix framework describes single diffractive data slightly better than the eikonal, regardless of the data employed.

Overall, despite the marginal difference in the functional form of the pomeron input in both schemes, we can deduce that the mechanism of unitarization significantly affects the fluctuations in the number of pomerons exchanged, and claim that the $U$-matrix scheme offers an efficient phenomenological approach to consider pomerons’ variability regardless of the energy range, while such fluctuations become significant only beyond specific high-energy thresholds in the other schemes, namely the eikonal and quasi-eikonal.

The intriguing energy shift with regards to the mean-variance relationship hints at a transition in the underlying dynamics of the hadronic scattering process with respect to the energy regime. Thus, in order to better comprehend this energy transition, we will examine the $f_2$ moment of the pomeron multiplicity distribution, i.e., the two-particle correlation parameter, measuring the correlation between pairs of pomerons, across various energies within both the eikonal and $U$-matrix frameworks. This parameter is defined as follows :

\begin{equation}
	f_{2} = < n (n - 1 )> - <n>^2 = D_{2}^2 - <n>
\end{equation} 
where $D_{2}$ is the dispersion: $D_{2}^{2} = <n^2> - <n>^2$. The two-particle correlation parameter has three possible values: negative, zero, and positive, in accordance with the multiplicity distributions that are narrower, equal to, or broader than a Poisson distribution.

\begin{figure}[htb]
	\centering
	\includegraphics[width=0.5\textwidth]{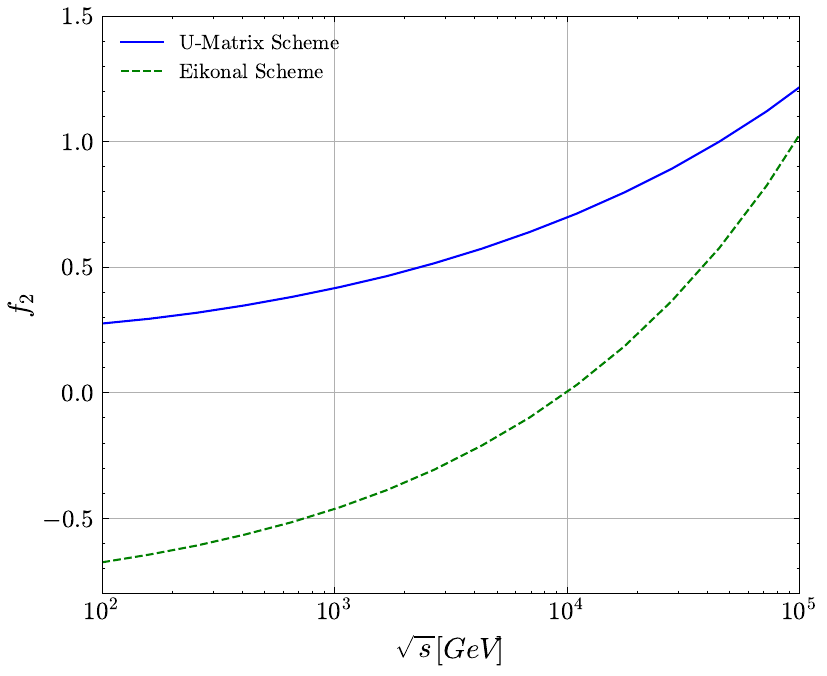}
	\caption{\label{fig:f2_pom_energy} The two-particle correlation parameter versus the interaction energy in the eikonal and $U$ Matrix case}
\end{figure}

By looking at Fig.~\ref{fig:f2_pom_energy}, we can vividly see in the eikonal case that the $f_2$ moment displays a changing behavior in variation with energy. Indeed, at energies below $10^4$ GeV, the $f_2$ moment is negative. This suggests that the pomeron multiplicity distribution is narrower than the Poisson distribution. This also implies that the particle production fluctuations are suppressed and the number of pomeron exchanges is distributed uniformly. At $10^4$ GeV, the value of the $f_2$ moment is zero, which is indicative of the alignment of both the pomeron multiplicity and the Poisson distributions. This entails independent and randomly occurring events. At energies above $10^4$ GeV, the $f_2$ moment becomes positive. This demonstrates that the pomeron multiplicity distribution is broader than the Poisson distribution, indicating an enhancement of pomeron fluctuations.

Turning now to the $U$-matrix case, Fig.~\ref{fig:f2_pom_energy} strikingly shows that the $f_2$ moment remains positive throughout various energy ranges, suggesting that the pomeron multiplicity distribution is constantly broader than the Poisson distribution and hence indicating an enhancement of pomeron fluctuations. These fluctuations become more noticeable with increasing energy and are remarkably constantly larger than those yielded in the eikonal case.

\begin{figure}[htb]
	\centering
	\includegraphics[width=0.5\textwidth]{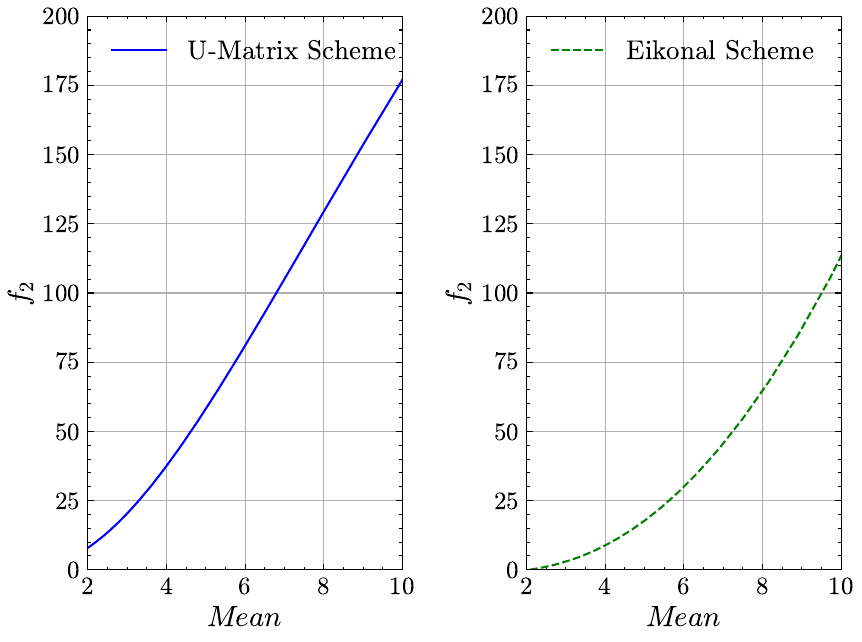}
	\caption{\label{fig:f2_mean} The two-pomeron correlation parameter versus the average pomeron in the eikonal and $U$ Matrix case}
\end{figure}

Fig.~\ref{fig:f2_mean} displays the behavior of the $f_2$ parameter with respect to the mean number of pomerons in the eikonal and $U$-matrix cases, specifically for energies exceeding $10^4$ GeV, where both schemes manifest pomeron fluctuations. It is apparent from this figure that both schemes yield amplified fluctuations. It also shows that the increasing mean number of pomerons with energy results in a rise in the correlation between pairs of pomerons. Nevertheless, it is worth noting that when comparing the two schemes, the correlation between pairs of pomerons rises considerably faster as the number of exchanged pomerons increases in the U-matrix scheme. This signifies that this latter incorporates enhanced correlations, indicating strong multi-pomeron dynamics.

It is evident from the sharper increase in $f_{2}$ in the $U$-matrix case that the two approaches handle high-energy hadronic interactions differently, with the $U$-matrix revealing more noticeable pomeron collective effects and non-linear pomeron exchanges.

These results strongly confirm our previous assertion that the $U$-matrix scheme naturally comprises pomeron statistical fluctuations that are distinct from those in eikonal-like schemes and further explain their impact on the properties of hadronic multi-particle production, in particular the unexpected overestimation of the fluctuations and correlations between final state particles with increasing energy in $pp$  collisions, as shown in \cite{Oueslati:2024awy}.

In view of the aforementioned results, we understand that the energy transition in the underpinning dynamics of the scattering process, which is only present in eikonal-like schemes, stems from a movement from a regime of suppressed pomeron fluctuations to a regime of enhanced fluctuations with increasing energy.

In light of the presence of enhanced pomeron fluctuations and a broader distribution within the $U$-matrix scheme, it is argued that the exchanged pomerons exhibit correlations which could be the result of collective effects, such as those stemming from the overlap of diffractive states that are in turn, emerging from the pomeron weights.

In an attempt to better comprehend these correlations, the higher-order moments of the pomeron multiplicity distribution were analyzed, as will be seen in the forthcoming subsection.
\subsection{Pomeron correlations}

In order to elucidate the nature of correlations among the pomerons exchanged in hadronic collisions, we examine the normalized factorial moments $F_{q} $  of the pomeron multiplicity distribution. In fact, the moment $F_{q}$ equals unity for all rank $q$ in the case of an independent exchange of pomerons. Nevertheless, $F_{q}$ is greater (less) than unity depending on whether the exchanged pomerons are correlated (anti-correlated) \cite{Dremin:1994fm}.
These moments are provided by

\begin{equation}
	F_{q}=\frac{1}{\langle n\rangle^{q}} \sum_{n=q}^{\infty}
	n(n-1)...(n-q+1)\,P_{n}\,,
	\label{eq:n_fatorial_m}
\end{equation}

where  $\langle n\rangle =\sum_{n} n  ~ W_{n}$ is the average multiplicity and $q$ is the rank of the moment. In the eikonal case, regardless of the energy scale, the normalized factorial moment remains equal to 1 for all ranks $q$, reflecting a Poisson distribution of the exchanged pomerons and indicating an uncorrelated distribution of events. In case of the $U$-matrix scheme, the moments are given by :

\begin{equation}
	E(X^{q})=p \,\ \mathrm{Li}_{-q}(1-p)    
\end{equation} where  $\mathrm{Li}_{-q}(1-p)$ is the polylogarithm function. So the normalized factorial moment of rank $q$ is given by : 

\begin{equation}
	\label{eq:n_fact_moment}
	\mathrm{F}_{q}(s) =  \frac{ p \,\ \mathrm{Li}_{-q}(1-p)}{
		(\frac{1-p}{p})^q} 
\end{equation} and $p$ defined by (\ref{eq:prob_succes}). 

\begin{figure}[htb]
	\centering
	\includegraphics[width=0.5\textwidth]{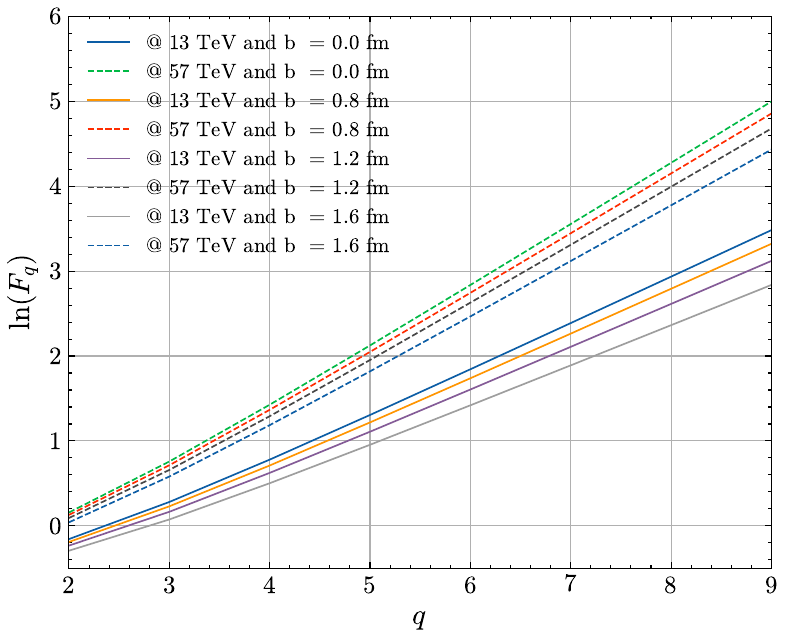}
	\caption{\label{fig:nfm_13_57_k} Normalized factorial moment of pomeron exchanges as a function of the rank q and for different impact parameter b values with the $U$ Matrix scheme}
\end{figure}

Fig.~\ref{fig:nfm_13_57_k} illustrates the behavior of the normalized factorial moments of pomerons exchanged in function of the rank q of pomerons and across various impact parameter b values within the framework of the $U$-matrix scheme, specifically at $13$ GeV and $57$ GeV. According to this figure, we can see that the normalized factorial moments $F_q$ exhibit a considerably increasing pattern with q at both energies, indicating stronger higher-order correlations, where pomeron correlations emerge from $q=3$ at $13$ GeV, with two pomerons showing anti-correlation. However, they appear starting from $q= 2$ at $57$ GeV, with all pomerons being positively correlated.

In both energies, when $b=0$ fm, $F_q$ is the highest suggesting that the strongest correlations occur in central collisions. Nevertheless, when b rises, we observe a decrease in $F_q$, indicating weaker correlations in less-central collisions because of a reduced interaction overlap. When comparing the two energies, $F_q$ values are consistently greater at $57$ GeV than at $13$ GeV for all pomerons exchanged and impact parameter b, showing that pomeron correlations become more intense as center-of-mass energy increases. In addition, at both energies, $F_q$ remains dependent on b, with correlations decreasing as b rises. Yet, it is significant to note that at $57$ GeV we have stronger correlations even at larger b. This highlights that increasing collision energy mediates the influence of the impact parameter on the correlation strength while simultaneously strengthening correlations and reducing the prevalence of anti-correlation effects.

On the whole, both the energy and the impact parameter b have a combined impact on the correlation between pomerons exchanged, indicating that these two parameters are not entirely independent when it comes to influencing the number of elementary interactions.

\begin{figure}[htb]
	\centering
	\includegraphics[width=0.48\textwidth]{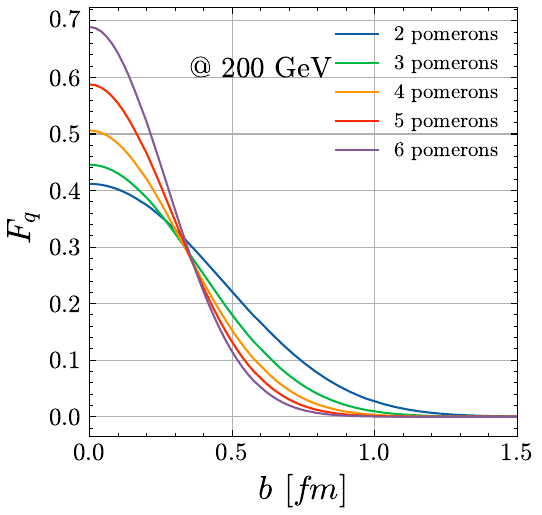}
	\caption{\label{fig:nfm_b_200} Impact parameter evolution of the normalized factorial moment at 200 GeV}
\end{figure}

\begin{figure}[htb]
	\centering
	\includegraphics[width=0.48\textwidth]{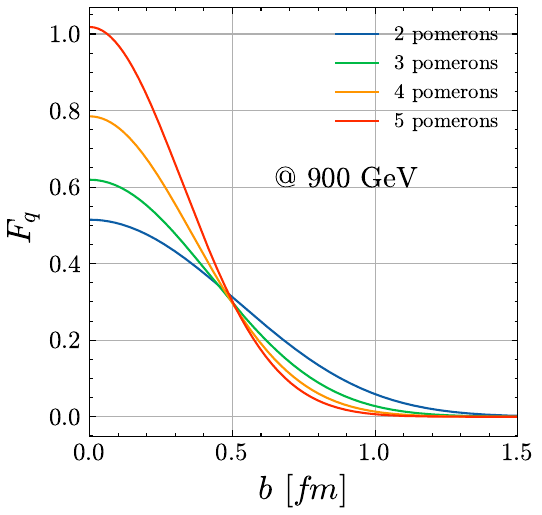}
	\caption{\label{fig:nfm_b_900} Impact parameter evolution of the normalized factorial moment at 900 GeV}
\end{figure}

\begin{figure}[htb]
	\centering
	\includegraphics[width=0.48\textwidth]{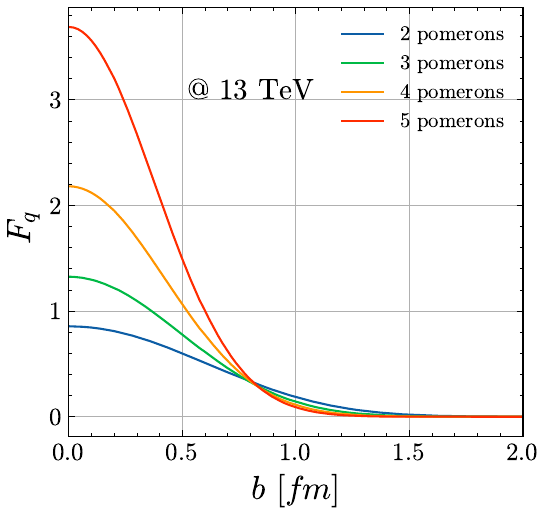}
	\caption{\label{fig:nfm_b_13} Impact parameter evolution of the normalized factorial moment at 13 TeV}
\end{figure}

\begin{figure}[htb]
	\centering
	\includegraphics[width=0.48\textwidth]{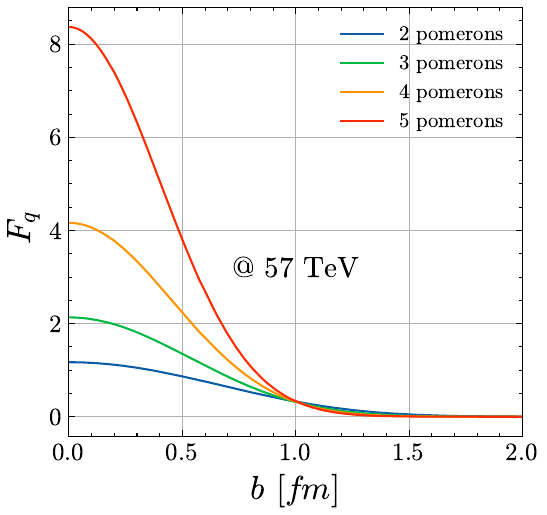}
	\caption{\label{fig:nfm_b_57} Impact parameter evolution of the normalized factorial moment at 57 TeV}
\end{figure}

To delve deeper in the interdependence between these two factors, we demonstrate in Fig.~\ref{fig:nfm_b_200}, \ref{fig:nfm_b_900}, \ref{fig:nfm_b_13} and \ref{fig:nfm_b_57} the impact parameter evolution of the normalised factorial moment for various energy scales and different pomerons exchanged. One intriguing trend revealed by this evolution is that for each energy, $F_q$ shows an inflection point at a specific value of b at which the correlation strength changes noticeably. Moreover, we can see that as energy increases, this inflection point moves to larger impact parameters, implying that the correlations between exchanged pomerons spread out more in transverse space. And, at extremely high energies, it reaches a value of about $1$ fm, where correlations are roughly zero. In this regard, we suggest that as energy grows, the spatial scope of the interactions becomes more significant, reflecting an interdependence between the energy scale and the transverse position of pomerons.

As a matter of fact, the aforementioned findings come in support of a previous suggestion that the $U$-matrix unitarization is probably incompatible with uncorrelated pomeron exchanges as there was no improvement in the description of the single diffractive data with a multi-channel model compared to a two-channel one  \cite{Oueslati:2023tjt}. To be more specific, \cite{Oueslati:2023tjt} a multi-channel model of high-energy hadron interactions was created by considering a full parton configuration space and using the $U$-matrix unitarization scheme. Moreover, the mean number of interactions between partons was assumed to be expressed as a product of the single-pomeron scattering amplitude, together with functions of the impact parameter and configurations. Furthermore, we assumed that the impact parameter had no effect on the distribution of parton configurations. Nevertheless, in the present study, the interdependence effect observed between s and b on the number of pomerons exchanged breaks down the rigorous validity of such factorization within the $U$-matrix scheme. Furthermore, aside from its ease of use as a workable framework, as with the eikonal approach, this factorization assumption is short of a compelling theoretical foundation. As a result, the parton distribution functions cannot be efficiently separated into longitudinal and transverse components. 

We can contend that the $U$-matrix approach offers a more suitable framework, as it provides a more accurate representation of the underlying elementary interactions, by taking into account the interdependence on all hadronic degrees of freedom. Moreover, correlations between pomerons allow the hadron's internal partonic structure to evolve dynamically during the scattering process. Thus, the $U$-matrix scheme does not require the hadrons to be frozen in their internal partonic configurations during the interaction, unlike the eikonal. Furthermore, by allowing for correlation between the exchanged pomerons, one can take into account the fact that the first pair of quark-antiquark strings is different from subsequent pairs, as the pomeron weights influence the dynamics of string pair formation. Consequently, the inconsistencies between the string model and the Gribov-Regge theory in hadronization models could be partially resolved within the $U$-matrix scheme.  

\subsection{Multiplicities in pp collisions}

Correlations between partons are known to occur as a result of the fundamental dynamics of partonic interactions as well as the spatial and momentum structure of the hadron. In fact, these correlations are critical in identifying the topological cross-sections for processes, such as double and triple parton scattering, which in turn influence the observed multiplicity patterns and cross-sections in hadronic collisions.

This subsection elucidates the role of the correlated multi-pomeron exchange in hadronic collisions by analyzing the multiplicity distribution of $ pp $ and $\bar{p} p $ collisions from the point of view of multi-parton interactions, as described by string models along with the Regge Phenomenology ( see the introduction section). According to this conjunction, a cut in the multi-pomeron exchange diagram is responsible for the hadrons yielded in the final state. More precisely, a cut of n pomerons results in $2n$ chains that connect to the partons of the initial hadrons. The number n, representing the pairs of simultaneously colliding partons from the different hadrons involved in the interaction, corresponds to the number of resulting showers. For instance, $n = 1$ translates into a single collision of one pair of partons emerging from the two colliding hadrons, which is ascribed to the Regge pole. $n= 2$ corresponds to a double collision of two pairs of partons from the different hadrons, which refers to the exchange of two pomerons, and so on.

In our analysis of the cross-section corresponding to the production of N secondary hadrons, $\sigma_N(s)$, diffraction processes were not considered so as to overlook long-distance correlations among particles within the same shower and the picture of the hadronic multi-particle production based on the Dual Parton Model (DPM) presented in \cite{Matinyan:1998ja} was followed mainly for comparison purposes. 

In fact,  in order to simplify our analysis, we assumed that the multiplicity distribution maintains its Poissonian character regardless of the energy scale, in spite of the known phenomenon of the violation of the KNO scaling \cite{KOBA1972317} which entails that as energy increases,  the hadronic multiplicity distribution broadens and deviates from a purely Poissonian nature : 

\begin{equation}
	P_n(N) = {\langle N_n\rangle^N\over N!}\; e^{-\langle N_n\rangle}
	\label{A1}
\end{equation}

where $\langle N_n \rangle  $ represents the mean number of particles produced in n showers and is taken proportional to the mean multiplicity for a single shower : 
\begin{equation}
	\langle N_n\rangle = n\langle N_1\rangle 
\end{equation} 
and the mean multiplicity for a single shower is modelled as : 
\begin{equation}
	\langle N_1\rangle =  a + b \; \ln{(s/
		s_0)}
	\label{mean_N1}
\end{equation} representing a logarithmic growth with centre-of-mass energy s, in agreement with experimental observations at low energy, with $a= -7.3$ and $b = 2.56$ from \cite{Matinyan:1998ja}. Thus, the total inelastic cross-section is constructed as a sum over contributions from n-shower events:

\begin{equation}
	\sigma_{in}(N,s) = \sigma_1P_1(N) + \sigma_2(P_2(N) +
	\sigma_3P_3(N) + \ldots , 
	\label{eq:sig_inela}
\end{equation}
where the pomerons weights $\sigma_n (\xi_n)$ are given after impact parameter  integration of  the cross-section for the production of $n$ showers  in both schemes  (\ref{eq:pom_topo_umat}), (\ref{eq:sig_n_eiko}) and   $\xi_n = \ln\left( {s\over s_0n^2}\right)$.  We retain three terms in \ref{eq:sig_inela} since the quadruple (or four-parton pair) collision has a small effect.

\begin{figure}[htb]
	\centering
	\includegraphics[width=0.48\textwidth]{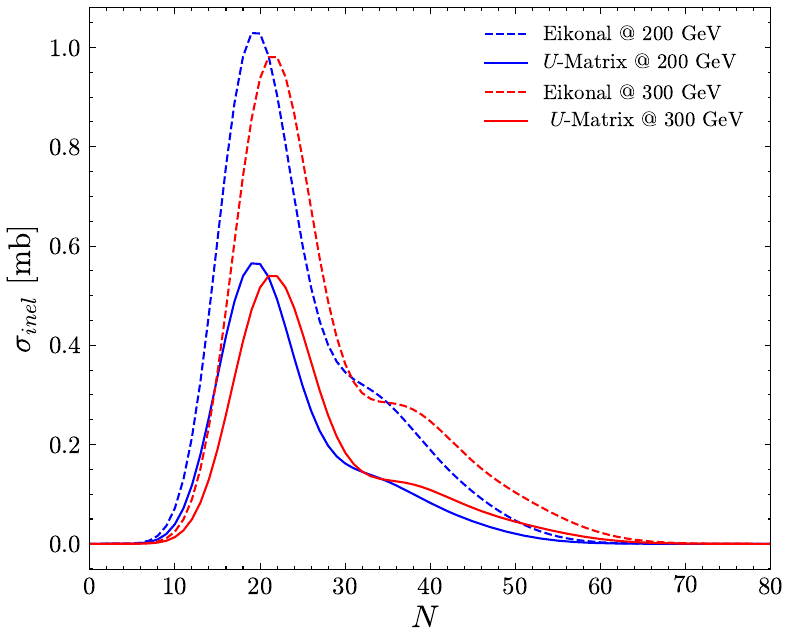}
	\caption{\label{fig:sigma_inel_low_energy} Topological cross sections $\sigma_{N}$ in the eikonal and $U$ Matrix approximation with exchanges of three effective soft Pomerons.}
\end{figure}

\begin{figure}[htb]
	\centering
	\includegraphics[width=0.48\textwidth]{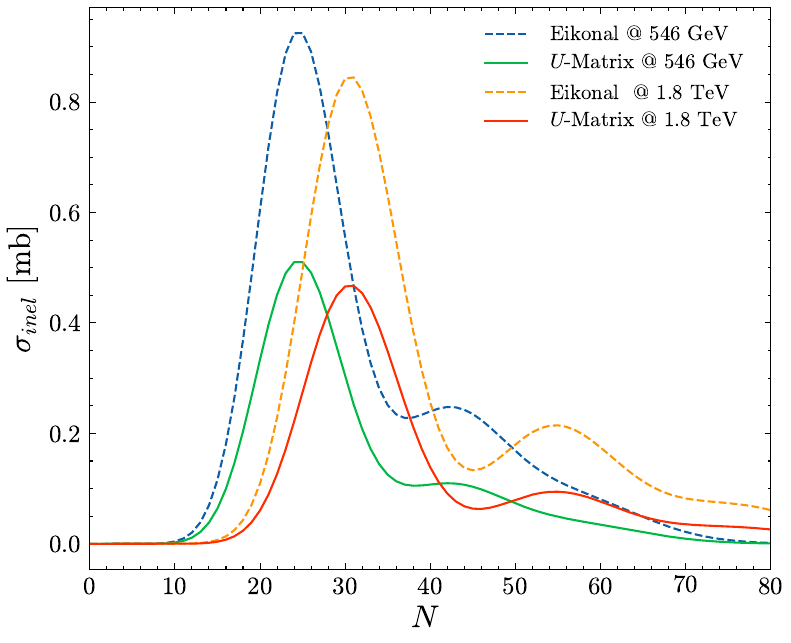}
	\caption{\label{fig:sigma_nch_high_energy} Topological cross sections $\sigma_{N}$ in the eikonal and $U$ Matrix approximation with exchanges of three effective soft Pomerons.}
\end{figure}

By looking at Fig.~\ref{fig:sigma_inel_low_energy} and Fig.~\ref{fig:sigma_nch_high_energy}, the shoulder, associated with the double collision, is clearly visible at low energies in the eikonal case, which is in concordance with the quasi-eikonal case \cite{Matinyan:1998ja}. Interestingly enough, in the $U$-matrix case, this second peak of $\sigma_N(s)$ is rather slightly resolved and tends to become broader and lower with increasing energy as opposed to the eikonal case. This behavior can be explained by the impact of correlated pomeron exchanges given that parton correlations enhance the probability of multi-parton collisions and hence re-distribute the contributions throughout the topological cross-sections.  Thus, in the $U$-matrix scheme, correlated pomeron exchanges play a key role in enhancing multi-parton collisions, particularly double parton collisions.

These outcomes suggest that pomeron exchanges in the $U$-matrix framework embed a probability of concurrently finding partons with momentum fractions $x_1$ and $x_2$ within the hadron, which is represented by a non-zero correlation function $F(x_1, x_2)$, emphasizing the deviation from independent multi-parton interactions. In this regard, one may wonder how the $U$-matrix approach could allow disentangling correlations in multiple parton interactions, specifically separating the effects of fractional momentum correlations from those of transverse coordinate correlations, which is beyond the remit of this paper.

It is worth noting that in \cite{Matinyan:1998ja}, the Dual Parton Model (DPM) was used to describe the double-parton collision, while solely taking the soft pomeron as the main component mediating the interactions between partons and comparing it to its hard counterpart. It has been argued that soft interactions are also responsible for the double inelastic parton collisions, which is confirmed by our result.

Moreover, given that the DPM is based on the quasi-eikonal scheme, we can infer that the incorporation of higher levels of diffractive production, resulting in an enhanced parton correlation, makes the $U$-matrix scheme a more reliable approach to the description of multi-parton collisions, compared to both the eikonal and quasi-eikonal schemes.

Owing that the soft interactions are also responsible for the double parton collisions, while a cut pomeron comprises contributions from both hard and soft processes,  this makes us wonder about its nature, particularly in relation to the hard-soft hadronic physics transition, and may pave the way for a unified description of high-energy hadronic collisions in the context of the $U$-matrix.


\section{CONCLUSION}

The chief purpose of the present paper was to understand the nature of the pomeron exchanges in hadronic interaction. Using a generalized representation of the unitarized hadronic elastic amplitude, the pomeron topological cross section was derived for both the $U$-matrix and eikonal schemes.  Our results have demonstrated that the mechanism of the multi-pomeron exchange summation is distinct in each scheme at many levels. To be more specific, it has been found that in the impact parameter space, the elementary interactions tend to occur at the core of the collision in the $U$- matrix case as opposed to the eikonal case.

In addition,  in both schemes, it has been shown that the pomeron topological cross-section for each higher-order exchanged pomeron increases with increasing energy and tends to reach a maximum then suddenly decreases at extreme energies, which marks the unitarity constraint. However, this unitarity constraint mark is reached at a somewhat higher energy for each pomeron exchanged in the $U$-matrix case. The disparity between the two schemes has also been shown concerning the curvature of this cross-section. More specifically, in the eikonal case, it considerably changed at energies beyond $10^4$ GeV. In the $U$-matrix case, no change was observed.

Furthermore, the pomeron multiplicity distribution has been computed. In the $U$-matrix case, the pomeron exchange is a random variable geometrically distributed, and the exchange of one pomeron has been demonstrated to enhance the probability of exchanging additional pomerons.

Moreover, in the $U$-matrix case, the number of pomerons exchanged has shown greater fluctuations than the average at each energy level, and this deviation becomes more significant as energy rises. It has been deduced that the $U$-matrix scheme intrinsically incorporates diffraction production into the multi-pomeron vertices, reflecting a larger pomerons’ variability regardless of the energy range, while such fluctuations become significant only beyond a specific high-energy threshold in the eikonal and quasi-eikonal schemes.

Furthermore, the pomeron exchange in the $U$-matrix scheme exhibits collective effects, as an increase in the number of exchanged pomerons leads to more pronounced higher-order pomeron correlations, which depend on both the energy and the impact parameter. This behavior contrasts with the independent pomeron exchange characteristic of the eikonal scheme.

Last but not least, the impact of pomeron weights on the proton-proton multiplicity distribution has been examined from the point of view of multi-parton interactions. The results have revealed that, in the $U$-matrix scheme,  correlated pomeron exchanges play a key role in enhancing multi-parton collisions, particularly double parton collisions.

We understand from the findings of this study that the pomeron distribution is fixed by the unitarization scheme chosen to satisfy the unitarity constraint, and that this choice cannot be arbitrary. 
While Monte Carlo event generators typically rely on eikonal models that require multi-parton correlations to be introduced through ad hoc mechanisms, rational unitarization schemes such as $U$-matrix, incorporate these correlations ab initio. Owing to the difficulty of computing the vertices of high-energy hadronic interactions from first principles, schemes like the $U$-matrix reflect the fundamental composite nature of hadrons much more naturally than eikonal-like approach. Nevertheless, it remains necessary to identify specific schemes  that strictly satisfy all mathematical and physical constraints. Constructing such a consistent, unified formalism thus constitutes a highly rich and stimulating perspective for future research.


\section*{Acknowledgments}

R.O. appreciates insightful discussions with Sergey Troshin and is grateful for his critical clarification regarding the physical interpretation of the $U$-matrix asymptotic inconsistency. Special thanks go to the computational resource provided by Consortium des Équipements de Calcul Intensif (CÉCI), funded by the Fonds de la Recherche Scientifique de Belgique (F.R.S.-FNRS) where a part of the computational work was carried out.

\bibliographystyle{apsrev4-2}
\bibliography{refs}

\end{document}